\documentstyle[preprint,aps,prb]{revtex}

\begin{document} 
 
\draft 
 
\title{Linear optical properties of one-dimensional Frenkel exciton systems
with intersite energy correlations} 

\author{V.\ A.\ Malyshev,$^{1,\dag}$ A.\ Rodr\'{\i}guez,$^{2}$ and 
F.\ Dom\'{\i}nguez-Adame$^{1}$} 

\address{$^{1}$GISC, Departamento de F\'{\i}sica de Materiales, 
Universidad Complutense, E-28040 Madrid, Spain\\
$^{2}$GISC, Departamento de Matem\'{a}tica Aplicada y Estad\'{\i}stica,
Universidad Polit\'{e}cnica, E-20840 Madrid, Spain}

\maketitle 
 
\begin{abstract} 
 
We analyze the effects of intersite energy correlations on the linear optical
properties of one-dimensional disordered Frenkel exciton systems. The
absorption line  width and the factor  of radiative rate enhancement are
studied as a function of the correlation length of the disorder. The absorption
line width monotonously approaches the seeding degree of disorder on increasing
the  correlation length. On the contrary, the factor  of radiative rate
enhancement  shows a non-monotonous trend, indicating a complicated scenario 
of the exciton localization in correlated systems.  The concept of coherently
bound molecules is exploited to explain the numerical results, showing good 
agreement with theory. Some recent experiments are discussed in the light of
the present theory.

\end{abstract} 
 
 
\section{Introduction} 

Theoretical studies of excited states of quasi one-dimensional (1D)
self-assembled molecular systems, like dipolar coupled molecules in 
J~aggregates,~\cite{Jelley36,Scheibe37} are often based on the well-known
Frenkel exciton model~\cite{Frenkel31} (see Ref.~\onlinecite{Davydov71}  for a
comprehensive review). The most remarkable optical properties of J~aggregates
are the narrowing of their absorption spectra 
(J~band)~\cite{Jelley36,Scheibe37} and the radiative rate 
enhancement~\cite{McRae58,DeBoer90} as compared to the monomers, both 
originating from the collectivization of the local molecular states due to the 
intermolecular dipolar coupling. The absorption band is narrowed by about  the
square root of the so-called {\em number of coherently bound molecules\/}, 
$N^{*}$, which, roughly speaking, is the number of molecules covered by  the
exciton wave function.\cite{Knapp84} The same number determines the factor of
radiative rate enhancement. $N^*$ is smaller than the physical length of an
aggregate since Anderson localization takes place  in 1D random
systems.~\cite{Mott61} It has been realized that diagonal  and off-diagonal
disorder are particularly important in determining the optical properties of
these systems because $N^{*}$ strongly depends  on the degree of
disorder.\cite{Knapp84,Fidder91}

Since the pioneering work by Knapp,\cite{Knapp84} theoretical studies of the
optical dynamics of Frenkel excitons in 1D random systems with intersite
correlations have attracted much attention.\cite{Knoester93,Adame95,Knoester96,%
Adame99,Adame00} Until recently, however, it was not possible to 
experimentally prove the occurrence of intersite correlations. Hence, most 
theoretical descriptions assumed that disorder was indeed uncorrelated. 
Knoester showed that nonlinear optical techniques are appropriate means to
determine both the degree of disorder and the correlation length of the
diagonal disorder.\cite{Knoester93} This approach was applied to molecular
aggregates of pseudo-isocyanine (PIC) by Durrant {\em et
al.\/}~\cite{Durrant94} and in aggregates of the dye
5,5',6,6'-tetrachloro-1,1'-diethyl-3,3'-di(4-sulfobuty)-benzimidazolocarbocianine
(TDBC) by Moll {\em et al.},~\cite{Moll95} which are known to easily form
J~aggregates.

One of the general consequences of the occurrence of intersite energy
correlations outlined in all the studies concerning the exciton absorption 
band is the increase of its width as compared to uncorrelated random systems.
\cite{Knapp84,Knoester93,Adame95,Knoester96,Adame99,Adame00} It was found  that
the absorption line width approaches the seeding degree of disorder on
increasing the correlation length  of the intersite energy disorder. To the
best of our knowledge, however, there was no a detailed study of the factor  of
radiative rate enhancement, carrying out information about the coherently bound
molecules. In a recent paper devoted to pairwise correlated diagonal disorder
(site energies appear at random to pairs of nearest-neighbor
sites),\cite{Adame99} we found the counterintuitive result that this factor
drops upon the occurrence of correlations. This means that excitons at the
bottom of the band become more localized after introducing pairwise correlations
while the underlying lattice is {\em less disordered\/}, in the sense that 
only half of the site energies are truly random variables.  Stronger
localization near the band edge upon introducing the correlations  in the
disorder
was also pointed out in the recent paper by Russ {\em et al.\/}~\cite{Russ98}

In this paper, a detailed study of 1D molecular systems with intersite energy
correlations is performed in order to uncover the effects of the short-range to
long-range correlation length crossover on the linear optical response of
Frenkel excitons. We show that the factor of radiative rate enhancement drops
upon increasing the energy correlation length whenever this length is smaller
than a given {\em characteristic\/} length. For larger energy correlation
lengths the behavior of the factor of radiative rate enhancement is the
opposite and increases on increasing the correlation length. We provide a
simple recipe for estimating this characteristic length. The absorption band
width presents a monotonous increase upon rising the energy correlation length
and approaches the width of the distribution of seeding disorder when the
correlation length exceeds the above mentioned characteristic length. We also
use the concept of coherently bound molecules to interpret numerical data.
Applying this concept in a self-consistent way, we find the scaling of the
factor of radiative rate enhancement and the width of exciton absorption line
with the energy correlation length and show that theoretical estimates
reasonably fit the results of numerical simulations. The remainder of the paper
is organized as follows. In the next Section~\ref{model}, we formulate the
model we will be dealing with. Section~\ref{Theory} deals with the theoretical
estimates which are then used in Sec.~\ref{Numerics} for explaining the resutls
of numerical simulations. Some recent experiments are discussed in the light 
of the present theory in Sec.~\ref{experiments}. Section~\ref{Summary}
summarizes the paper.

\section{Model of disordered system}
\label{model} 
 
We consider $N\gg 1$ optically active, two-level molecules forming a regular 1D
lattice with spacing unity. To build up our correlated disordered lattice with
intersite energy correlation length $N_c$, we create $S \equiv N/N_c$
consecutive segments with equal transition energies $\epsilon_s$ within each
segment. The set $\{\epsilon_s\}_{s=1}^{S}$ are statistically independent
Gaussian variables with probability distribution 
\begin{equation}
p(\epsilon_s) = \left(\frac{1}{2\pi\sigma^2}\right)^{1/2} 
\exp \left( -\,{\epsilon_s^2\over 2\sigma^2}\right),
\label{p}
\end{equation}
and $\langle\epsilon_s\epsilon_{s^{\prime}}\rangle = \sigma^2
\delta_{ss^{\prime}}$, where brackets denote the average over the join energy
distribution
\begin{equation}
P(\{\epsilon_s\}) = \prod_{s=1}^{S} p(\epsilon_s), 
\label{P}
\end{equation} 
and $\sigma$ is the seeding degree of disorder.

For our present purposes, we will neglect all thermal degrees of freedom
(electron-phonon coupling and local lattice distortions) and assume that
disorder originates from a static offset of the on-site energies (diagonal
disorder). The effective tight-binding Hamiltonian for the 1D Frenkel-exciton
problem in the random lattice with correlated disorder can be then written as
follows
\begin{equation} 
{\cal H}=\sum_{s=1}^{S}\>\epsilon_s \sum_{n=1}^{N_c} |n+N_c(s-1)\rangle 
\langle n+N_c(s-1)| + \sum_{n,m=1}^N \> J_{nm} \> |n\rangle \langle m|,
\label{Hamiltonian} 
\end{equation} 
where $|\nu\rangle$ is the state vector of the $\nu\,$th molecule and $J_{nm} =
J_{|n-m|}$ are the hopping integrals, which are assumed to be negative, as it
takes place for J~aggregates, and not subjected to disorder.

We are interested in the absorption line shape and its width as well as in the
factor of enhancement of the exciton radiative rate relative to the monomer 
spontaneous emission rate. The absorption line shape is calculated as
\begin{equation}
I(E)={1\over N}\bigg\langle\sum_{j=1}^{N}\>\mu_{j}^{2}\> 
\delta \left(E-E_j \right) \bigg\rangle,
\label{oas}
\end{equation}
Here we assume that the system size is much smaller than the optical
wavelength. The index $j$ runs over all eigenstates of the 
Hamiltonian~(\ref{Hamiltonian}) and $E_j$ denotes their corresponding
eigenenergies. The oscillator  strength of the $j\,$th eigenstate with
components $a_{j}^{(n)}$ is given  by 
\begin{equation}
\mu_{j}^{2} \equiv \left( \sum_{n=1}^{N}\>a_{j}^{(n)}\right)^{2},
\label{mu}
\end{equation}
where the dipole moment of an isolated monomer is taken to be unity. 

The factor of radiative rate enhancement is defined through the average
oscillator strength per state at energy $E$~\cite{Fidder91,Schreiber82}
\begin{equation}
\mu_{\mathrm av}^2(E)= {I(E)\over \rho(E)},
\label{aos}
\end{equation}
where 
\begin{equation}
\rho(E)={1\over N}\bigg\langle\sum_{j=1}^{N}\>\delta \left(E-E_j \right) 
\bigg\rangle
\label{dos}
\end{equation}
is the normalized density of states. We take $\mathrm{max}\{\mu_{\mathrm
av}^2(E)\}$ as a measure for the enhancement of the exciton radiative
rate~\cite{Fidder91} or, in other words, for the number of coherently 
bound molecules $N^*$.~\cite{Knapp84} In what follows, the above magnitudes 
will be determined both analytically and numerically as a function of the 
correlation length $N_c$.

\section{Theoretical estimates}
\label{Theory}

Having presented our model, we now describe the method we have used to estimate
the factor of radiative rate enhancement $N^*$ and the linear absorption band
width $\sigma_1$. To this end, we rewrite the Hamiltonian (\ref{Hamiltonian})
in the excitonic representation using the eigenfunctions of the unperturbed
Hamiltonian (with $\epsilon_s=0$ for $s=1,\ldots,S$). For the sake of
simplicity, we assume here periodic boundary conditions. The Bloch plane waves
are then the exact eigenfunctions of (\ref{Hamiltonian}) in the absence of
disorder:
\begin{equation}
|K\rangle = {1\over\sqrt N} \sum_{n=1}^N
e^{iKn}|n\rangle,
\label{K}
\end{equation}
where $K=2\pi k/N$ belongs to the first Brillouin zone ($0 \le k < N$). In
the $K$-representation, the Hamiltonian (\ref{Hamiltonian}) reads
\begin{mathletters}
\label{1}
\begin{equation}
{\cal H} = \sum_K E_K |K\rangle \langle K| + \sum_{K,K^\prime}
\Delta_{KK^\prime} |K\rangle \langle K^\prime|, 
\label{Hex}
\end{equation}
where $E_K$ is the unperturbed exciton energy 
\begin{equation}
E_K = \sum_{n=1}^N \> J_{n}e^{iKn}, 
\label{E_K}
\end{equation}
and $\Delta_{KK^\prime}$ is the intermode scattering matrix 
\begin{equation}
\Delta_{KK^\prime} = {1\over N}\sum_{s=1}^S \epsilon_{s}
\sum_{n=1}^{N_c} \> e^{i(K - K^\prime)[ n + N_c (s-1)]}.
\label{DeltaKK'} 
\end{equation}
\end{mathletters}
\noindent  
The role of $\Delta_{KK^\prime}$ is two-fold. The off-diagonal elements
$\Delta_{KK^\prime}$ ($K\ne K^\prime$) mix the exciton states. If the typical
fluctuation of $\Delta_{KK^\prime}$ does not exceed the energy difference
between the states $K$ and $K^\prime$ (the so called perturbative limit) then
the energy shifts of the corresponding states, given by the diagonal elements
$\Delta_{KK}$, is the main effect of the disorder and result in inhomogeneous
broadening of exciton levels for an ensemble of chains. The typical fluctuation
of $\Delta_{KK}$ has a direct relationship with the inhomogeneous width of the
exciton state $\big|K\big\rangle$.

For nonperturbative magnitudes of the disorder, off-diagonal elements of the
scattering matrix, $\Delta_{KK^\prime}$ ($K\ne K^\prime$), mix the exciton
states, resulting in their localization on chain segments of a typical size
smaller than the chain length and subsequently affecting the exciton optical
response.  Recall that, for a perfect circular chain, only the state
$|K=0\rangle$ is coupled to  the light and carries the entire exciton
oscillator strength, which is then $N$ times larger than that for an isolated
molecule. Being mixed with other (nonradiative) states ($K\neq 0$), the
radiative state loses a part of the oscillator strength due to its spreading
over the nonradiative ones. Thus, the number of coherently bound molecules $N^*
< N$ arises as the enhancement factor of the oscillator strength of the
localized exciton states.~\cite{Knapp84} Accordingly, the inhomogeneous width
of the optical exciton line will also be subjected to renormalization since $N$
should be replaced by $N^*$.~\cite{Knapp84}

\subsection{Perturbative treatment}
\label{MNE}

Assuming the perturbative limit (a quantitative condition of its validity 
will be given below), let us gain insight into the magnitude of the typical 
fluctuation of the scattering matrix $\Delta_{KK^\prime}$. To this end, one
should calculate its second moment using the joint distribution~(\ref{P}) of
the seeding fluctuations (being expressed trough linear combinations of 
Gaussian variables $\epsilon_s$ with zero mean, the $\Delta_{KK^\prime}$ are
also Gaussian variables with $\langle\Delta_{KK^\prime}\rangle=0$). The
magnitude of our interest will be $\sigma_{KK^\prime}^2 \equiv \Big\langle
|\Delta_{KK^\prime}|^2\Big\rangle$. It is an easy task to arrive at
\begin{equation}
\sigma^2_{KK^{\prime}} = {\sigma^2\over NN_c}\>
{\sin^2[(K-K^\prime)N_c/2] \over \sin^2[(K-K^\prime)/2]}.
\label{variance_gen}
\end{equation}
In the limits $K,K^\prime \ll 1$ and $(K-K^\prime)N_c \ll 1$, which are 
of major importance from the viewpoint of optical properties, 
Eq.~(\ref{variance_gen}) reduces to 
\begin{equation}
\sigma^2_{KK^{\prime}} = \sigma^2 \> {N_c\over N}.
\label{variance_part}
\end{equation}
This expression generalizes the result we obtained previously for pairwise
correlated disorder ($N_c = 2$)~\cite{Adame99} to the case  of an arbitrary
energy correlation length $N_c$. As follows from Eq.~(\ref{variance_part}), the
motiona narrowing effect ($N^{-1}$ scaling) is weakened here by the factor $N_c$
and is determined now by the chain length $N$ counted in units of the energy
correlation length $N_c$ or, in other words, by the number of correlation
segments $S=N/N_c$. The motional narrowing effect  completely disappears at 
$N_c=N$. Similar conclusions were drawn in 
Refs.~\onlinecite{Knapp84,Fidder91,Knoester93,Adame95}.

As it was already mentioned above, the state $|K=0\rangle$ carries the entire 
oscillator strength of the system so that the optical absorption spectrum
presents an isolated Gaussian peak centered around the eigenenergy $E_{K=0}$
and it is characterized by its standard deviation 
\begin{equation}
\sigma_{1} = \sigma \sqrt{N_c\over N}.
\label{sigma_1}
\end{equation}
This result holds whenever the state $K=0$ does not become mixed with the 
nearest one $|K_1\rangle$, where $K_1=2\pi/N$. The mixing is governed  by the
off-diagonal element $\Delta_{K_10}$. Hence, one should compare 
$\sigma_{K_10}$ with the energy difference $E_{K_1}-E_0$. The perturbative 
approach is still applicable provided $\sigma_{K_10} < E_{K_1}-E_0$ and  fails
otherwise. Thus, the equality $\sigma_{K_10} = E_{K_1}-E_0$ determines a value
of $\sigma$ (for fixed both chain length $N$  and energy correlation length 
$N_c$)
which separates the ranges of perturbative and nonperturbative magnitudes of
disorder. Taking $E_K$ in the nearest-neighbor approximation
\begin{equation}
E_K = - 2J\cos(K) \simeq -2J + JK^2,
\label{E_NN}
\end{equation}
where $J$ is the nearest-neighbor coupling and $N$ is assumed to be large, 
one gets
\begin{equation}
\sigma = {4\pi^2J\over (N^3N_c)^{1/2}},
\label{threshold}
\end{equation}
which determines the crossover between the perturbative and non-perturbative
limits.

\subsection{Nonperturbative limit}
\label{CBM}

At nonperturbative magnitudes of disorder, not all molecules of the chain 
contribute to the optical spectra of the whole ensemble, but only a portion of
them being a function of the degree of disorder $\sigma/J$.~\cite{Knapp84}  The
reason is the localization of the excitonic states arising from disorder.
Therefore, the number of coherently bound molecules, $N^{*}$, (averaged number
of molecules covered by optically active localized exciton states) should
replace the number of molecules in the system, $N$. In
Ref.~\onlinecite{Victor93}, a simple rule for estimating $N^*$ was formulated
whenever the disorder is uncorrelated. It exploits the findings that i) the
lowest localized exciton states can be classified in several groups of states
(two or sometimes three), each one localized on a certain chain segment of a
typical size $N^*$, ii) each segment does not overlap with the
others~\cite{Victor93} (see also Refs.~\onlinecite{Malyshev95,Shimizu98} and
Sec.~\ref{Numerics} for more details) and, what is most important, the states
of each group have the energy structure similar to that for an homogeneous
chain of size $N^*$, i.e., approximately given by Eq.~(\ref{E_NN}) with $N$
replaced by $N^*$.

The rule proposed in Ref.~\onlinecite{Victor93} for estimating $N^*$ consists
simply of applying the perturbative criterion (\ref{threshold}) to a typical
localization segment, i.e., substituting $N$ by $N^*$ in Eq.~(\ref{threshold})
and considering now $N^*$ as an unknown parameter.  As was shown in
Refs.~\onlinecite{Victor93,Malyshev95}, this approach works surprisingly well
in fitting the numerical data concerning the optical response of 1D Frenkel
excitons. In doing so, one arrives at the following estimate of the number 
of coherently bound molecules 
\begin{equation}
N^{*}= {1\over N_c^{1/3}}\left( {4\pi^2J\over \sigma} \right)^{2/3}.
\label{N^*}
\end{equation}
According to Ref.~\onlinecite{Knapp84}, the standard deviation of the exciton  
absorption spectrum $\sigma_1$ can be estimated by Eq.~(\ref{sigma_1}),  
again replacing $N$ by $N^{*}$, that gives us
\begin{equation}
\sigma_{1}=N_c^{2/3}\, \sigma \left({\sigma \over 4\pi^2J}\right)^{1/3}.
\label{sigma^*}
\end{equation}
The $N_c$ scaling laws of $N^*$ and $\sigma_1$ are the main results  of
the present paper. Note that $\sigma$-scaling of $N^*$ and $\sigma_1$   was
previously reported in Refs.~\onlinecite{Fidder91,Victor93} and
\onlinecite{Fidder91,Schreiber82,Victor93,Kohler89,Tilgner90,Boukahil90},
respectively. We stress that the number of coherently bound molecules drops
while the standard deviation of the exciton absorption spectrum rises upon 
increasing the intersite energy correlations, meaning stronger scattering of
the exciton in a more correlated system. This somewhat unexpected result
stems from the fact that motion narrowing effect is partially suppressed 
as intersite energy correlations occurs, yielding a smaller reduction of 
the seeding disorder as compared to the case of uncorrelated disorder.

\subsection{Short- to long-range correlations crossover}
\label{Crossover}

The concepts presented in the previous subsection as well as the scaling 
laws~(\ref{N^*}) and~(\ref{sigma^*}) are valid within the range  $N_c < N^*$.
However, both numbers, $N_c$ and $N^*$, will approach  each other as the
correlation length $N_c$ rises and, starting from a certain  value of $N_c$ we
will obtain $N^* \approx N_c$. We now provide arguments how to estimate the
magnitude of $N_c$ at such a crossover. As soon as $N^*$ and $N_c$ approach
each other, there is no reason for the reduction of disorder due to the
motional
narrowing since the energy of any segment of size $N_c$ is fixed, i.e, does not
fluctuate. Thus, the  typical energy mismatch between two adjacent segments is
of the order of  $\sigma$. On the other hand, the fact that $N^* \approx N_c$
means that  the coupling between adjacent segments is of the order or smaller 
than their energy separation $\sigma$.  From this it follows that the 
condition we are looking for is precisely that under which two adjacent 
segments of size $N_c$ start to be decoupled of each other. 

In line with this reasoning, let us consider the two segment problem, taking
for the sake of simplicity the corresponding Hamiltonian in the 
nearest-neighbor approximation
\begin{eqnarray}
{\cal H} & = & \sum_{\alpha=a,b} \left[\sum_{n=1}^{N_c} 
\epsilon_\alpha |\alpha, n\rangle\langle n,\alpha| 
- J\sum_{n=1}^{N_c-1} \Big(|\alpha, n\rangle \langle  n+1,\alpha| 
+ |\alpha, n+1\rangle\langle n,\alpha|\Big)\right]\nonumber\\
\nonumber\\
& - & J \Big(|a,N_c\rangle \langle 1,b| + |b,1\rangle \langle N_c,a|\Big),
\label{h}
\end{eqnarray}
where $a$ and $b$ are the segment labels and the second term describes the
coupling of the segments. Making use of the excitonic transformation 
\begin{equation}
|\alpha,n\rangle = \left({2\over N_c+1}\right)^{1/2}
\sum_K \sin (Kn)\> |\alpha,K\rangle \ , 
\label{n}
\end{equation}
where $K=\pi k/(N_c+1)$ and $k$ ranges from 1 to $N_c$, one passes 
from~(\ref{h}) to
\begin{mathletters}
\label{2}
\begin{equation}
{\cal H} = \sum_{\alpha=a,b}\sum_K (\epsilon_{\alpha}+E_K) 
|\alpha,K\rangle \langle K,\alpha| 
+ \sum_{K,K^\prime} ({\cal H}_{ab})_{KK^\prime} 
|a,K\rangle \langle K^\prime,b| \ . 
\label{h_ex}   
\end{equation}
Here, $E_K$ is given by Eq.~(\ref{E_NN}) and the interaction Hamiltonian
reads
\begin{equation}
({\cal H}_{ab})_{KK^\prime} = (-1)^{k+1}{2J\over N_c + 1} 
\sin(K) \sin(K^\prime)\ . 
\label{h_ab} 
\end{equation}
\end{mathletters}
\noindent
The interaction Hamiltonian $({\cal H}_{ab})_{KK^\prime}$ couples excitonic
modes belonging to different segments and thus leads to their  spreading 
over both segments. Recall that, from the viewpoint of the optical response, 
mixing of the lowest modes with $K_1 = K^\prime_1 = \pi/(N_c  + 1)$ is of 
major importance. We aim to elucidate for which segment size  $N_c$ this 
coupling can be considered as a perturbation.  Remembering that the typical 
energy mismatch between the segments is $\sim \sigma$
and taking $|({\cal H}_{ab})_{K_1K^\prime_1}|=\sigma$, one arrives at the 
following estimate of the size we are looking for
\begin{equation}
N_c^* +1 = \left({2\pi^2J \over \sigma}\right)^{1/3},
\label{N_c^*}
\end{equation}
where the condition $N_c \gg 1$ was used. 

Summarizing, the typical energy separation of two adjacent  segments starts to
dominate over their coupling at $N_c > N_c^*$ so that the segments can   be
considered as decoupled of each other. As a consequence, the localization
length of the lowest exciton mode will be the same that the energy correlation
length $N_c$.

\section{Discussion of the numerical results} 
\label{Numerics}

In order to calculate the quantities of interest defined by 
Eqs.~(\ref{oas})-(\ref{dos}) we replaced $\delta(E-E_j)$ by the Heaviside
step-function $(1/R)\theta (|E-E_j| - R)$, where $R$ is the resolution, taken
to be 0.05 in all the calculations. For energy units, $J=1$ was chosen. The
standard deviation $\sigma$ was also fixed to be 0.2. The only varied parameter
was the energy correlation length $N_c$.

We diagonalized the Hamiltonian (\ref{Hamiltonian}) for open chains of size
$N=SN_c$, where for a given $N_c$, the number $S$ was chosen such that $N$
would be the closest one to 250. The number of randomly generated systems was 
$1000$
for each value of $N_c$. We do not present here the exciton absorption spectra
since they are rather standard ones, showing the characteristic asymmetry and
red shift due to disorder discussed in detail for uncorrelated disorder in
Refs.~\onlinecite{Fidder91,Schreiber82,Klafter} and for the correlated case in 
Refs.~\onlinecite{Knapp84,Adame95,Adame99}. In Fig.~\ref{fig1}, the standard
deviation of the absorption spectra $\sigma_1$ is shown as a function of the
energy correlation length $N_c$. To accurately determine the standard deviation
$\sigma_{1}$ we fitted the low-energy side of the spectra using Gaussians. One
can see from Fig.~\ref{fig1} that $\sigma_1$ monotonously increases as $N_c$
rises and finally approaches the value $0.2~(=\sigma)$ for the case at hand.
The crossover, occurring approximately at $N_c^* = 5$, is in  good agreement
with the estimate $N_c^* \approx 4$ given by Eq.~(\ref{N_c^*}). 

Figure~\ref{fig2} shows the behavior of the factor of radiative rate
enhancement $\mathrm{max}\{\mu_{\mathrm av}^2(E)\}$ as a function of $N_c$.
First, it drops upon increasing $N_c$ and then it grows for $N_c > 4$,  also in
full correspondence with our theoretical prediction done  in Sec.~\ref{Theory}.
Strong fluctuations observed in Fig.~\ref{fig2} result from the fluctuations 
in the density of states: the latter appears in the denominator  in the
definition of $\mathrm{max}\{\mu_{\mathrm av}^2(E)\}$ [see Eq.~(\ref{aos})]. We
do not present in Fig.~\ref{fig2}  how  the quantity
$\mathrm{max}\{\mu_{\mathrm av}^2(E)\}$ behaves for  $N_c > 15$ for several
reasons. First, as soon as the segments of correlated energies start to be 
decoupled  of
each other (the case for $N_c > 4 $, see Fig.~\ref{fig3}b and~\ref{fig3}c),
the  oscillator strength per state $\mu_{\mathrm av}^2(E)$ behaves
monotonously,  i.e., has no maximum (see Appendix~\ref{apendice}), thus proving
to be an  inconvenient measure of the extension of the optically active exciton
states.  Second, it is evident that, for decoupled segments, the localization
length  $N^*$ is equal to the energy correlation length $N_c$.
Figure~\ref{fig3}  confirms the last statement. There, we presented typical
realizations of  the eight lowest eigenfunctions of the Hamiltonian
(\ref{Hamiltonian}) for  three values of the energy correlation length $N_c=4$,
$10$, and $40$.  They demonstrate that the energy segments actually turn out to
be  decoupled of each other as the segment size rises. Indeed, for $N_c = 4$, 
the typical extension of the eigenfunctions, $N^*$, of the lowest excitonic 
states is more than $10$ (the oscillator strengths $\mu^2$ of nodeless  states
carry this information). Note that it sufficiently exceeds the  energy
correlation length. For larger value, $N_c=10$, both magnitudes  approach  each
other,  although still keeping the relation $N^*>N_c$.  Besides, in the last
case, the eigenfunctions tend to be grouped on different  segments of size
$\approx N_c$. Finally, for the largest presented value of  $N_c (=40)$, the
eigenfunctions $\varphi_{Kn}$,  oscillator strengths  $\mu_K^2$ and
eigenenergies $E_K$ within each group follow fairly well the  corresponding
formulae for an isolated segment of size $N_c$:
\begin{mathletters}
\label{3}
\begin{equation} 
\varphi_{Kn} =  \left({2 \over N_c+1}\right)^{1/2} 
\sin \left( Kn \right), 
\end{equation}
\begin{equation} 
\mu_K^2 ={1 \over 2(N_c+1)}\Big[1 - (-1)^k \Big]^2 
\cot^2 \left({K \over 2}\right), 
\end{equation}
\begin{equation} 
{\cal E}_K = \epsilon + E_K, 
\end{equation}
\end{mathletters}
\noindent
where $\epsilon$ is the site energy of the molecules within each segment.  Note
that such a structure (called hidden in Ref.~\onlinecite{Victor93}) of the
exciton low-energy spectrum is also present for $N_c =4$ but scaled in average
by the number $N^*$ instead of $N_c$.

\section{Discussion of the experimental results} 
\label{experiments}

In the recent experimental works by Durrant {\em et al.\/}~\cite{Durrant94} 
and by Moll {\em et al.}~\cite{Moll95} it was reported on the presence of
strong correlations in diagonal disorder in molecular aggregates of PIC and
TDBC, respectively. Moreover, it was found in both studies that the energy
correlation length $N_c$ is of the order of the localization
length~\cite{Durrant94} and even larger.~\cite{Moll95} This limit strongly 
suggests that almost no motion narrowing effect has to be 
detected.~\cite{Knoester93,Knoester96} Nevertheless, the observed J~band  width
in both cases was definitely motionally narrowed. The possible ordering of the
first solvatation shell in the presence of highly ordered aggregate during the
freezing process and the subsequent reduction of the local disorder was given
as a possible explanation for the J~band narrowing. 

It seems to be interesting to estimate the critical length $N_c^*$ for   the
above aggregates on the basis of the experimental data presented in 
Refs.~\onlinecite{Durrant94,Moll95}. Taking $J=600\,$cm$^{-1}$ and
$\sigma=13.5\,$cm$^{-1}$ for J~aggregates of PIC,~\cite{Durrant94} one gets
$N_c^* \approx 9$ what is consistent with the relation $N_c \simeq N^*$  found
by Durrant {\em et al.\/}~\cite{Durrant94} For J~aggregates of TDBC
($J=850\,$cm$^{-1}$ and $\sigma=67\,$cm$^{-1}$), one arrives at $N_c^*\approx
6$. Thus, the experimental estimation of the energy correlation length $N_c$ to
be of the order of  several hundreds molecules obtained in 
Ref.~\onlinecite{Moll95} contradicts, in our opinion, the magnitude of
localization  length $N^*\approx 40$ extracted from the spectroscopic data.

\section{Summary and concluding remarks} 
\label{Summary} 

In summary, we have studied the effects of intersite energy correlations on the
characteristics of linear absorption spectrum corresponding to a weakly
localized 1D Frenkel exciton. We have found that the factor of radiative rate
enhancement shows a nonmonotonous behavior upon rising $N_c$, first decreasing
as $N_c^{-1/3}$ and then growing linearly. The crossover occurs when the
typical coupling of segments is smaller than the typical separation between
their energy levels, being of the order of the width of the seeding Gaussian
energy distribution $\sigma$. The linear trend indicates that the segments
become independent one from the other when $N_c > (2\pi^2 J/\sigma)^{1/3}$. The
standard deviation of the absorption line monotonously  increases as
$N_c^{2/3}$ for low energy correlation length and saturates when $N_c >
(2\pi^2J/\sigma)^{1/3}$, namely when the localization length $N^*$  coincides
with the energy correlation length $N_c$. Our results lead us to a better
understanding of recent experimental results in PIC and TDBC molecular
aggregates.

\acknowledgments 

This work was supported by CAM under project No.~07N/0034/1998.  V.~A.~M.\@
thanks UCM for the support under project {\em Sab\'{a}ticos Complutense\/}.

\appendix

\section{Evaluation of $\mu_{\mathrm \lowercase{av}}^2(E)$ for decoupled
segments}  
\label{apendice}

Let us calculate $\mu_{\mathrm av}^{-2}(E) = \rho(E)/I(E)$, assuming the
energy segments to be decoupled of each other. Under such
an assumption, one can rewrite $\rho(E)$ and $I(E)$ as follows
\begin{mathletters}
\label{A1}
\begin{equation}
\rho(E)={1\over N}\bigg\langle \sum_{s=1}^S \sum_{k=1}^{N_c}\>
\delta \left(E-\epsilon_s - E_K \right) 
\bigg\rangle,
\label{A1dos}
\end{equation}
\begin{equation}
I(E)={1\over N}\bigg\langle \sum_{s=1}^S \sum_{k=1}^{N_c}\mu_{K}^{2}
\>\delta \left(E-\epsilon_s -E_K \right) \bigg\rangle,
\label{A1oas}
\end{equation}
\end{mathletters}
\noindent
where $E_K = - 2J\cos[\pi k/(N+1)]$, $k=1,2,\ldots,N_c$ and 
$\{\epsilon_s\}_{s=1}^S$ is a stochastic Gaussian energy offset distributed
according to Eq.~(\ref{P}). Making use of the integral representation of the
$\delta$-function $\delta(x)=(1/2\pi) \int_{-\infty}^{\infty}dt\exp(ixt)$,
it is easy to carry out the average in Eqs.~(\ref{A1dos}) and~(\ref{A1oas})
explicitly to arrive at
\begin{mathletters}
\label{A2}
\begin{equation}
\rho(E)={1\over N_c}\sum_{k=1}^{N_c}\> 
\left(\frac{1}{2\pi\sigma^2}\right)^{1/2} 
\exp \left[-\,{(E-E_K)^2\over 2\sigma^2}\right],
\label{A2dos}
\end{equation}
\begin{equation}
I(E)= {1\over N_c}\sum_{k=1}^{N_c}\> \mu_K^2
\left(\frac{1}{2\pi\sigma^2}\right)^{1/2} 
\exp \left[-\,{(E-E_K)^2\over 2\sigma^2}\right].
\label{A2oas}
\end{equation}
\end{mathletters}
\noindent
Accounting further for the fact that the lowest exciton state (with $K_1 =
\pi/[N_c+1])$ carries almost the entire oscillator strength ($\mu_{K_1}^2/
\sum_K \mu_K^2 = 8/\pi^2$, namely $8\%$) and thus keeping in Eq.~({\ref{A2oas})
only the first term, we finally get
\begin{equation}
\mu_{\mathrm av}^{-2}(E) = {\pi^2 \over 8N_c}
\left\{ 1 + \sum_{k=2}^{N_c} \exp \left[{E_K - E_{K_1}\over\sigma^2}
\left(E - {E_K+E_{K_1}\over 2}\right)\right]\right\} \ .
\label{fin}
\end{equation}
{}From this it follows that $\mu_{\mathrm av}^{-2}(E)$  monotonously rises 
from the value $\pi^2/8N_c$ to infinity, when passing from $E=-\infty$ to
$E=\infty$. As a consequence, $\mu_{\mathrm av}^{2}(E)$ monotonously  
decreases from the value $8N_c/\pi^2$ to zero. 
It should be pointed out  that the characteristic energy scale of the
exponentials in Eq.~(\ref{fin}) 
is determined by the ratio $(E_K-E_{K_1})/ \sigma^2$ which goes down
as the  segment size $N_c$ rises. This means that close to the band bottom,
which is precisely the energy region we are interested in, the contribution
of the exponentials in Eq.~(\ref{fin}) may be large. Then, the magnitudes
of $\mu_{\mathrm av}^{2}(E)$ obtained from numerical simulations will 
strongly differ from the asymptotic value $8N_c/\pi^2$.
Due to the peculiarity outlined above,  the oscillator strength per state 
$\mu_{\mathrm av}^{2}(E)$ appears to  be an inconvenient measure for the 
extension of the optically active exciton  states for the case of larger 
energy correlation length, when the energy segments are decoupled one from 
the other.

\begin{figure} 
\caption{Standard deviation  of the exciton absorption spectra $\sigma_{1}$
versus the energy correlation length $N_c$. The degree of seeding disorder is
$\sigma=0.2$.} 
\label{fig1} 
\end{figure} 
 
\begin{figure} 
\caption{Factor of the radiative rate enhancement, $\mathrm{max}\{\mu_{\mathrm
av}^2(E)\}$, versus the energy correlation length $N_c$. The degree of seeding
disorder is $\sigma=0.2$.} 
\label{fig2} 
\end{figure} 

\begin{figure} 
\caption{Typical realizations of the lower exciton eigenmodes for three values
of the energy correlation length a) $N_c = 4$, b) $10$, and c) $40$.} 
\label{fig3} 
\end{figure} 

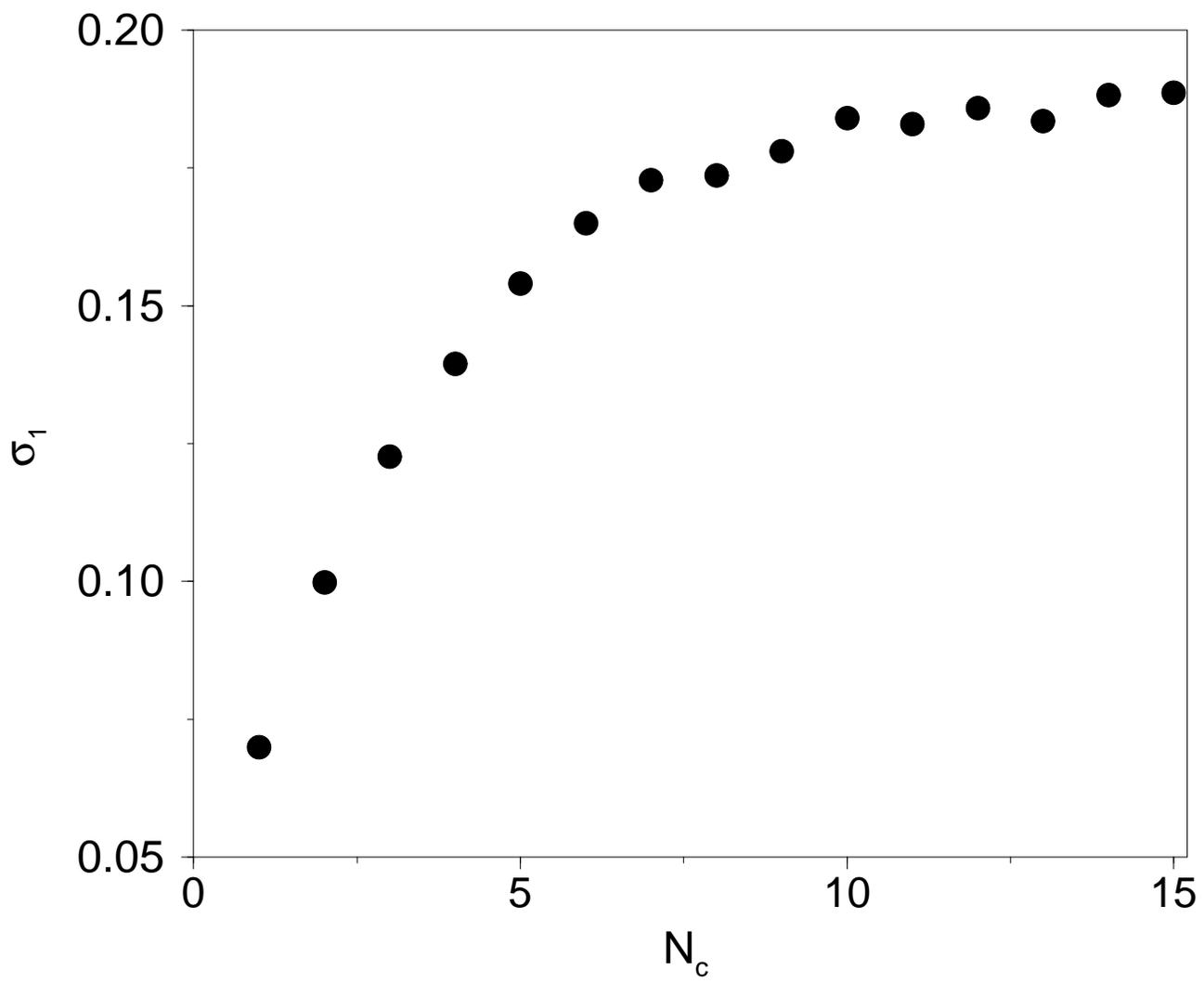
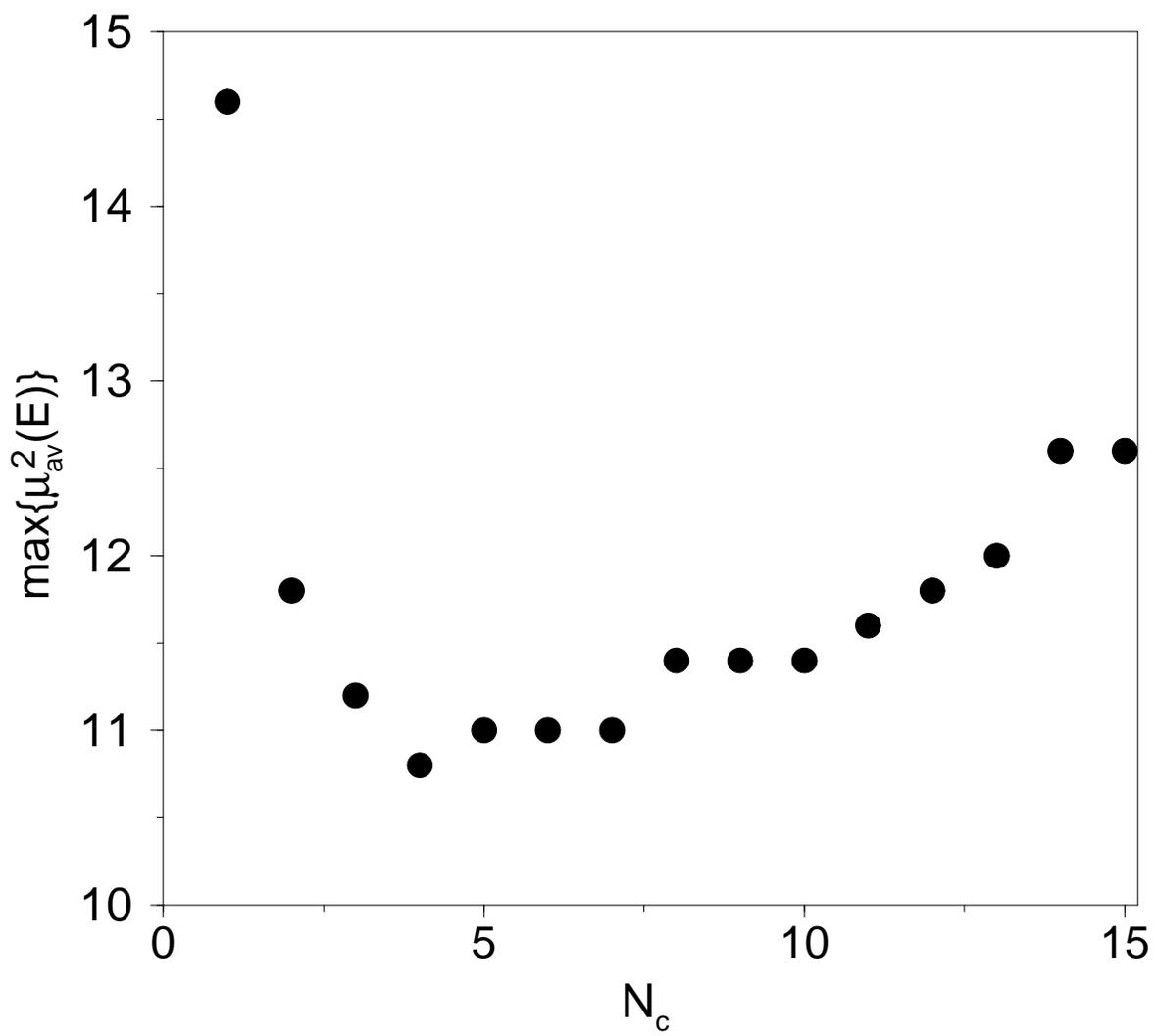
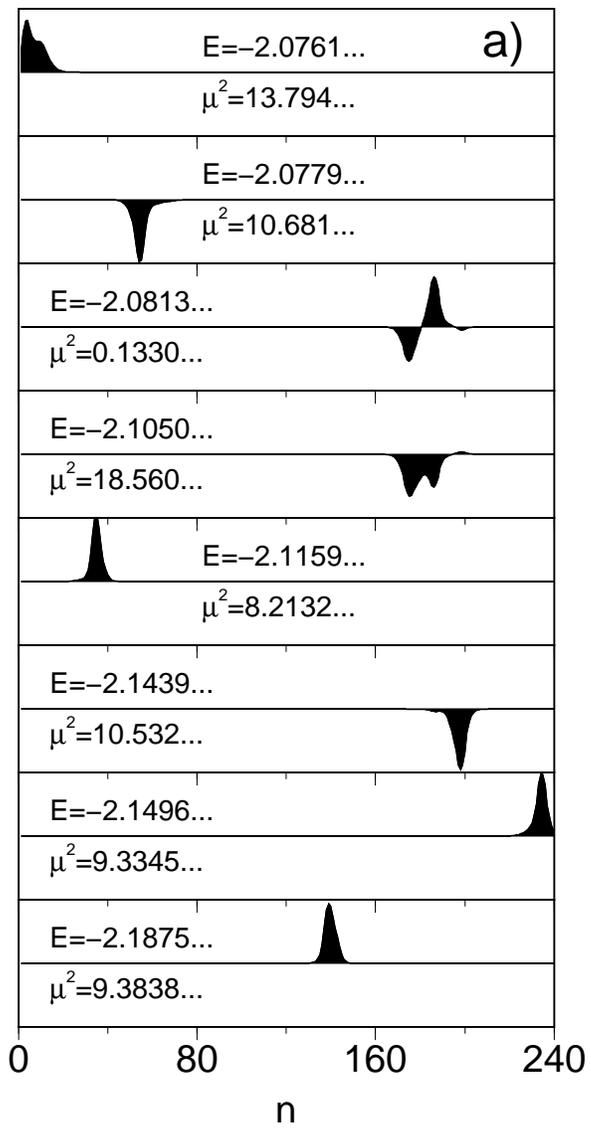
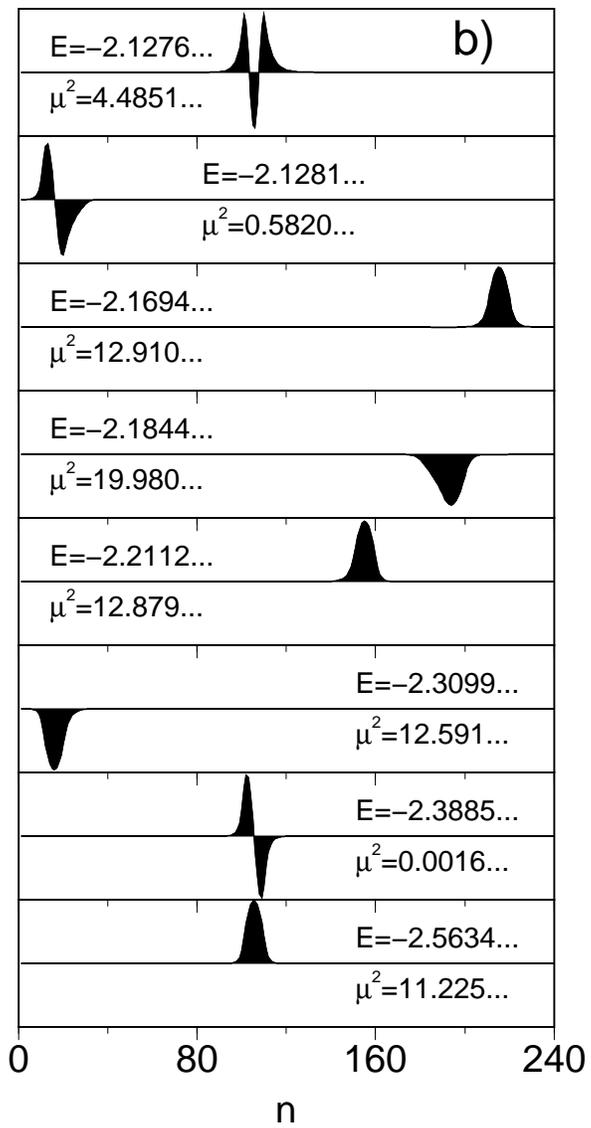
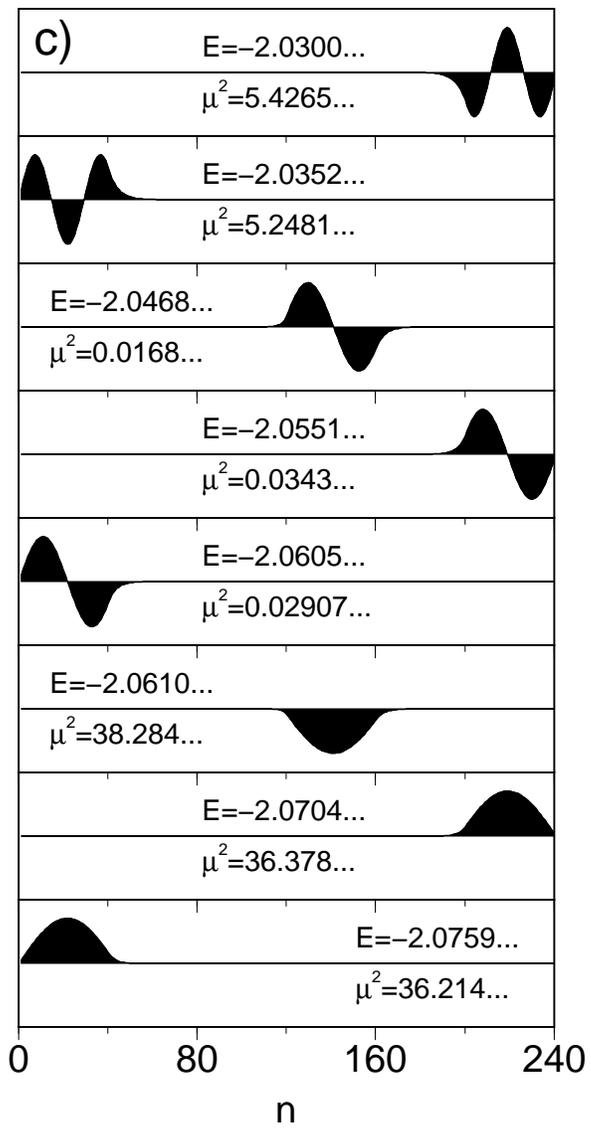
\end{document}